\documentclass[a4paper,11pt]{article}
\usepackage{jinstpub} 
\usepackage{lineno}

\usepackage{lipsum}

\proceeding{New Frontiers in Lepton Flavor\\
Pisa, Italy\\
15--17 May 2023}

\title{Looking for an axion in a haystack of muons}

\author{A. Gurgone}
\collaboration[c]{on behalf of the \textsc{McMule} team}
\affiliation{Dipartimento di Fisica, Universit\`a di Pavia,\\ INFN, Sezione di Pavia,\\ Via A. Bassi 6, 27100 Pavia, Italy}

\emailAdd{andrea.gurgone01@ateneopv.it}

\abstract{
The search for axion-like particles $X$ in muon decays is an excellent opportunity for the MEG II and Mu3e experiments to extend their horizons beyond ${\mu^+ \to e^+ \gamma}$ and  ${\mu^+ \to e^+ e^- e^+}$.
A suitable process for both experiments is the two-body decay ${\mu^+ \to e^+ X}$, whose only signature is a monochromatic peak close to the kinematic endpoint of the positron energy spectrum of the ${\mu^+ \to e^+ \nu_e \bar\nu_\mu}$ background.
The hunt for such an elusive signal in a vast amount of irreducible background requires extremely accurate theoretical predictions to be implemented in a Monte Carlo event generator.
This work presents a new state-of-the-art computation of ${\mu^+ \to e^+ \nu_e \bar\nu_\mu}$ for polarised muons, accomplished with the \textsc{McMule} framework. 
The calculation includes next-to-next-leading order QED corrections and logarithmically enhanced terms at even higher orders.
The results are also used to estimate the sensitivity of both experiments on the branching ratio of ${\mu^+ \to e^+ X}$, in order to evaluate the impact of the theoretical error.
}

\keywords{Simulation methods and programs; Radiation calculations}

\begin{document}
\maketitle
\flushbottom

\section{Introduction}

The search for charged Lepton Flavour Violation~(cLFV) in muon decays is a sensitive probe to test the Standard Model~(SM) at the intensity frontier.
The MEG II~\cite{MEGII:2018kmf} and Mu3e~\cite{Arndt:2020obb} experiments at the Paul Scherrer Institut~(PSI) are designed to detect ${\mu^+ \to e^+ \gamma}$ and  ${\mu^+ \to e^+ e^- e^+}$ with an unprecedented accuracy. 
In addition, both experiments appear to be suitable for muon decays producing a new light neutral boson $X$, such as an axion-like particle (ALP)~\cite{Calibbi:2020jvd}.
In this regard, a viable channel for both experiments is given by the two-body decay ${\mu^+ \to e^+ X}$, where the ALP is supposed to escape undetected.
Since the muon decays at rest, the only signature of ${\mu^+ \to e^+ X}$ is a monochromatic peak close to the kinematic endpoint of the positron energy spectrum of the ${\mu^+ \to e^+ \nu_e \bar\nu_\mu}$ background. 
The hunt for such an elusive signal in a vast amount of irreducible background requires extremely accurate theoretical predictions, especially at the energy endpoint, where the radiative corrections are logarithmically enhanced by the emission of soft photons.

\section{Theoretical predictions}

\begin{figure}[t]
\centering
\includegraphics[width=.85\textwidth]{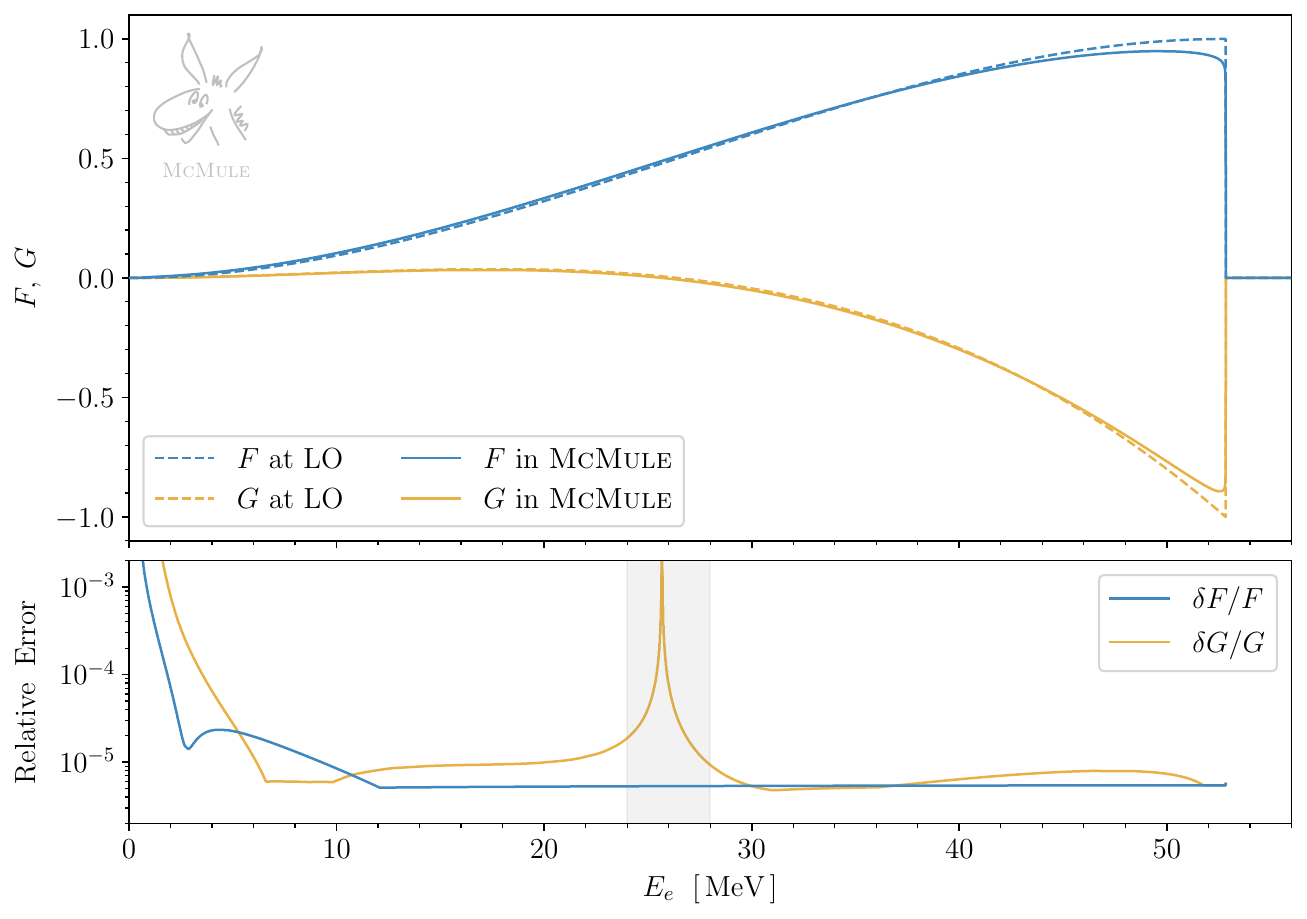}
\caption{\label{fig:background} The $F$ and $G$ functions for ${\mu^+\to e^+\nu_e \bar\nu_\mu}$ in \textsc{McMule}, compared with the leading order prediction. 
The positron energy spectrum in the total solid angle is simply $2F$.
The lower panel shows the resulting theoretical error, based on the estimation of the missing perturbative contributions. 
The value of $\delta G / G$ is artificially enhanced in the grey region because $G \simeq 0$ for $E_e \simeq m_\mu /4$.}
\end{figure}

The theoretical predictions for both ${\mu^+ \to e^+ X}$ and ${\mu^+ \to e^+ \nu_e \bar\nu_\mu}$ have been computed with the fully differential Monte Carlo code \textsc{McMule}~\cite{Banerjee:2020rww}. 
Assuming inclusive photons, the differential decay width of a positive muon with polarisation $|P_\mu| \leq 1$ can be written as 
\begin{linenomath*}
\begin{equation}
\label{eq:fg}
\frac{m_\mu}{2} \frac{\textrm{d}^2\Gamma}{\textrm{d}E_e\,\textrm{d}\hspace{-1pt}\cos\theta_e} = \frac{G_F^2 \, m_\mu^5}{192 \, \pi^3}\, \Big[ F(E_e) + P_\mu \cos\theta_e \, G(E_e) \Big] \, ,
\end{equation}
\end{linenomath*}
where $E_e$ denotes the positron energy, $\theta_e$ the angle between the positron momentum and the muon polarisation, $G_F$ the Fermi constant, and $m_\mu$ the muon mass. 
The two energy functions $F$ and $G$ fully characterise the positron kinematics and can be computed perturbatively.
In \textsc{McMule} the signal ${\mu^+\to e^+X}$ is implemented assuming a generic cLFV coupling and including the QED corrections at next-to-leading order~(NLO).
The background ${\mu^+\to e^+\nu_e \bar\nu_\mu}$ includes the leading weak and hadronic contributions, the full QED corrections at next-to-next-leading order~(NNLO), and collinear logarithmic terms at even higher orders.
At the endpoint of the positron energy spectrum, the emission of soft photons results in large logarithms that are resummed up to next-to-next-to-leading logarithmic~(NNLL) accuracy. 
The details of both calculations with an evaluation of the theoretical error can be found in~\cite{Banerjee:2022nbr}.
The result for ${\mu^+\to e^+\nu_e \bar\nu_\mu}$ is shown in Figure~\ref{fig:background}.

\section{Expected sensitivity}

\begin{figure}[t]
\centering
\includegraphics[width=.97\textwidth]{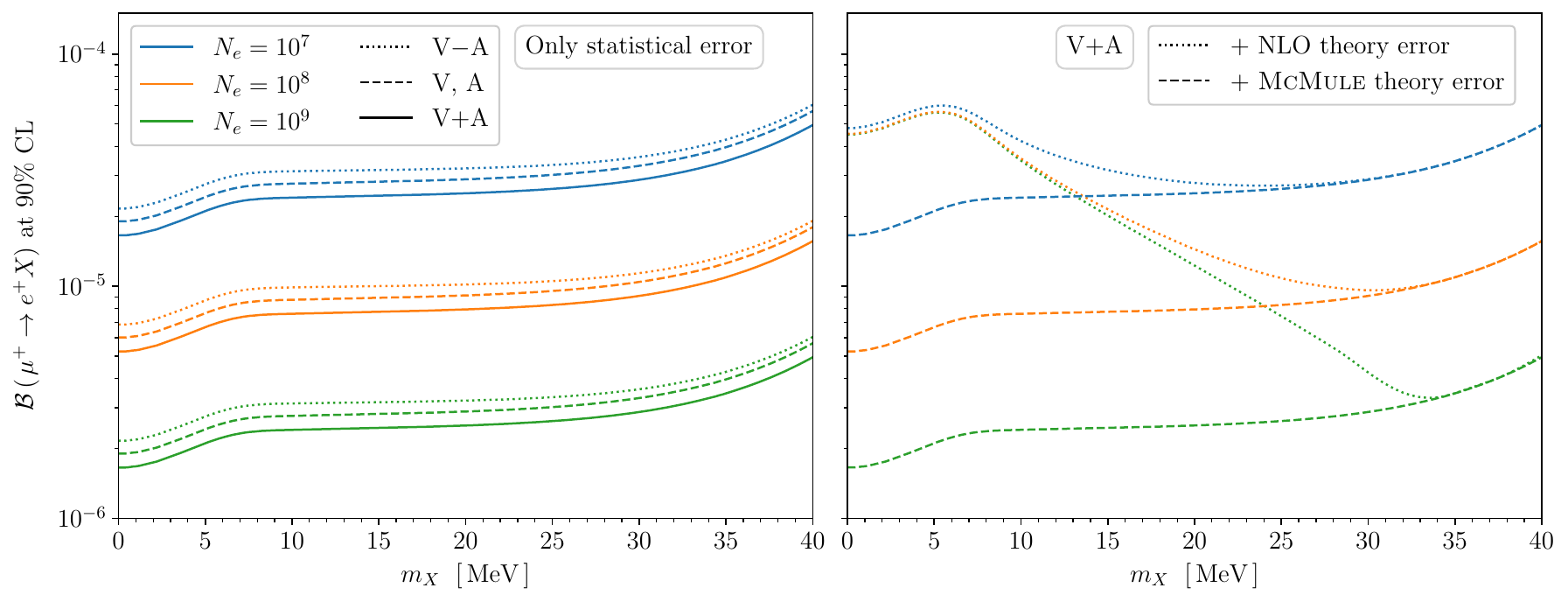}
\caption{\label{fig:sensitivity} Estimated sensitivity on the branching ratio of ${\mu^+ \to e^+ X}$ for the MEG II experiment.  
The left panel includes only the statistical error contribution, for different assumptions of ALP couplings and number of positron events.
The right panel shows the effect of adding the theoretical error for a $V+A$ coupling. 
Using only a NLO prediction for the background, the sensitivity would have been limited by theory, not statistics.
}
\end{figure}

The expected positron energy spectrum $\mathcal{P}$ in both experiments can be obtained by convoluting the theoretical spectrum $\mathcal{H}$ of both decays as  
\begin{align}\label{eq:pdf}
\mathcal{P}(E_e) &=\int dE'\, \big[ \mathcal{H}(E'_e) \times 
\mathcal{A}(E'_e) \times \mathcal{S}(E_e, E'_e) \big]
\equiv 
\left( \mathcal{H} \times \mathcal{A} \right) \otimes \mathcal{S}
\, ,
\end{align}
where $\mathcal{A}$ denotes the energy acceptance function and $\mathcal{S}$ the detector response function, which can be parametrised according to the detector simulations~\cite{MEGII:2018kmf,Arndt:2020obb}.
An indication of the expected sensitivity on the branching ratio of ${\mu^+\to e^+X}$ can be estimated by applying the cut-and-count analysis described in~\cite{Banerjee:2022nbr} to $\mathcal{P}$.
Assuming a massless ALP with a $V+A$ coupling to leptons, the results suggest a branching ratio sensitivity of about $2\cdot10^{-6}$ for MEG~II and $10^{-8}$ for Mu3e, which can rely on more statistics.
These predictions can be compared with the upper limit of $2.5\cdot10^{-6}$ at 90\% C.L. measured in~\cite{Jodidio:1986mz}.
More detailed results for MEG~II, including the effect of adding the theoretical error, are reported in Figure~\ref{fig:sensitivity}.
However, since a positive offset in the reconstruction of the positron energy results in an excess of events at the endpoint, a rigorous control of the systematic errors is required to avoid signal biases for $m_X \simeq 0$.
The development of new calibration tools for the positron energy is therefore essential for a reliable search in both experiments.

\section{Conclusions}
The search for ALPs in cLFV muon decays is an excellent opportunity for MEG~II and Mu3e to complement their main searches with additional competitive channels.
The hunt for the elusive two-body decay ${\mu^+ \to e^+ X}$ requires a new state-of-the-art computation of the SM background ${\mu^+\to e^+\nu_e \bar\nu_\mu}$, which has been achieved with the \textsc{McMule} framework.
The results have been used to implement a new positron event generator in the experimental software~\cite{Gurgone:2022nzy}.
This is the first important step towards more detailed studies to assess the feasibility of this challenging but intriguing search.

\acknowledgments
The author is grateful to Pulak Banerjee, Antonio M. Coutinho, Tim Engel, Angela Papa, Patrick Schwendimann, Adrian Signer, and Yannick Ulrich for their collaboration.


\bibliographystyle{JHEP}
\bibliography{biblio.bib}

\end{document}